\documentclass[12pt,thmsa]{article}
\usepackage{amssymb}

\usepackage{sw20aip}

\input{tcilatex}
\begin{document}

\title{Relation of the Bell inequalities with quantum logic, hidden variables and
information theory}
\author{Emilio Santos \and Departamento de F\'{i}sica. Universidad de Cantabria.
Santander. Spain}
\maketitle

\begin{abstract}
I review the relation of the Bell inequalities - characteristic of
(classical) probabilities defined on Boolean logics - with noncontextual and
local hidden variables theories of quantum mechanics and with quantum
information.
\end{abstract}

\smallskip Quantum mechanics looks radically different from all classical
theories of physics. Is that difference fundamental or is it due to our
present lack of understanding of quantum mechanics?. (That our understanding
is not good enough is widely recognized \cite{laloe}.) Several approaches
have been followed in the attempt to answer the question. We comment here
briefly on three of them, namely quantum logics, hidden variables and
information theory, showing the relation of each one of these with the Bell
inequalities.

\section{Quantum logic and quantum probability}

According to Birkhoff and von Neumann \cite{BN} the difference between
quantum and classical theories is radical because it appears at the most
fundamental level, the logic. The elements of a logic are the propositions
which, using the language of physics, are observables having the possible
values 1 (the proposition is true) or 0 (false). Some pairs of propositions
are related by the \textit{implication (A implies B }if $B$ is true whenever 
$A$ is true). This binary relation endowes the logic with the mathematical
structure of a partially ordered set (''poset''). Another binary relation
associates every proposition with its negation (for each propostion A there
exist another one, A', which is true if and only if the first is false).
This makes the poset \textit{orthocomplemented. }The internal operations
''meet'' and ''join'' endowes the poset with a richer structure making it an
orthocomplemented \textit{lattice}. Finally it is assumed that there exist
the sure proposition $I$, always true, and the absurd proposition, always
false, which makes the lattice \textit{complete}. From now on any complete
and orthocomplemented lattice will be called a \textit{logic}. Classical
logic is a \textit{distributive} lattice and it is called a \textit{Boolean
algebra}.

In the view of Birkhoff and von Neumann the structure of quantum logic may
be derived from the correspondence between propositions and \textit{%
projection operators (} which we shall call projectors in the following.)
Accordingly these authors postulated that the proposition associated to the
projector $\widehat{P}$ is true (or false), for a physical system in a given
state, if the state-vector $\mid \Psi \rangle $ is an eigenvector of $%
\widehat{P}$ $\left( \text{ or }\widehat{I}-\widehat{P}\right) .$ This
assumption gives rise to a \textit{trivalent} logic where propositions may
be, in addition to true or false, also undefined (which happens if $\mid
\Psi \rangle $ is neither an eigenvector of $\widehat{P}$ nor an eigenvector
of $\widehat{I}-\widehat{P}.)$ As projectors are associated to closed
subspaces of the Hilbert space, quantum logic has the mathematical structure
of the set of closed subspaces.

From these assumptions it is straightforward to define the fundamental
relation of order (or \textit{implication}) of propositions. We say that,
for two propositions $A$ and $B$ we have $A\leq B$ ( or $A\Rightarrow B$) if
the subspace associated to B contains that associated to A. Hence the binary
operations ''meet'', $\curlywedge ,$ and ''join'' ,$\curlyvee ,$ may be
defined in a natural form and it follows that the propositions form a 
\textit{lattice}. The lattice is \textit{orthocomplemented} (the subspaces
assiciated to the proposition A and its negation A' are orthogonal) and 
\textit{complete} (there exist the sure proposition, $I$, corresponding to
the whole Hilbert space and its negation, $\Phi ,$ corresponding to the null
vector). Up to here everything is similar to what happens in classical
logic. But the quantum lattice is not \textit{distributive (Boolean}) at a
difference with the classical one. As a conclusion the authors claimed that
the non-Boolean character of the lattice of propositions is the essential
chracteristic of quantum theory. The details may be seen in the original
article \cite{BN}.

In the 66 years elapsed since the work of Birkhoff and von Neumann many
articles and several books have been devoted to the subject of quantum logic
(see e.g. the book of Hooker\cite{H}), in many cases starting from different
definitions of quantum propositions. Also some criticisms have aroused in
the sense that ''quantum logic'' is not a true logic, but just a
propositional calculus. Indeed in an ''actual'' logic the relations amongst
proposition like $A$ $\Rightarrow B$ or $A\curlywedge B=C$ should be also
considered propositions, which is not necesarily the case in a propositional
calculus. But the commented approach to the logic of quantum mechanics is
still widely used.

In any logic (orthocomplemented and complete lattice) it is straightforward
to define a probability distribution (or ''state''):

\begin{definition}
If $\mathcal{L}$ is a logic, a probability distribution is a mapping p:$%
\mathcal{L\rightarrow }\left[ 0,1\right] $ with the axioms

1) p($\Phi )=0,p(I)=1,$ where $\Phi $ ($I$ ) is the absurd (sure)
proposition,

2) If \{$A_{i}$\} is a sequence such that $A_{i}\leq A_{j}^{\prime }$ , $A$'
being the negation of $A$, for all pairs $i\neq j$ , then $%
\sum_{i}p(A_{i})=p\left( \curlyvee A_{i}\right) ,$

3) For any sequence \{A$_{i}$\}, p(A$_{i}$) = 1 $\forall i\Rightarrow
p\left( \curlywedge A_{i}\right) =1,$
\end{definition}

Thus from quantum logic, as defined by Birkhoff and von Neumann, we get
quantum probability, whilst the classical, Boolean, logic provides the
standard probability theory. Indeed the above axioms are simply a
generalization of the axioms of probability as stated by Kolmogorov.

\section{The Bell inequalities}

A discrimination between classical and quantum probability is provided by
the Bell inequalities, derived as follows \cite{S}. For any two proposition
A, B $\in \mathcal{L}$ we may define a function, d(A,B), by 
\begin{equation}
d(A,B)=p\left( A\curlyvee B\right) -p\left( A\curlywedge B\right) .
\label{23}
\end{equation}
That function has the properties 
\begin{equation}
0\leq d(A,B)\leq 1,d(A,A)=0,d(A,A^{\prime })=1,  \label{24}
\end{equation}
and provides some measure of the ''distance'' between two propositions in a
given state (probability distribution). The function is called a \textit{%
metric (pseudometric) }if the following additional property holds (does not
hold) true 
\begin{equation}
d(A,B)=0\Rightarrow A=B,  \label{25}
\end{equation}
but this property is not very relevant for our purposes. More important are
the following \textit{triangle inequalities}, which are (are not generally) 
\textit{fulfilled if the lattice is }(is not) \textit{Boolean } 
\begin{equation}
\left| d(A,B)-d(A,C)\right| \leq d(B,C)\leq d(A,B)+d(A,C).  \label{26}
\end{equation}
As the Boolean character provides the essential difference between classical
and quantum theories, according to Birkhoff and von Neumann \cite{BN}, we
see that the triangle inequalites $\left( \ref{26}\right) $ give a criterium
to distinguish both theories. These inequalities are closely related to the
Bell inequalities as shown in the following \cite{S}, although in
mathematical theory of probability the inequalities $\left( \ref{26}\right) $
were known well before Bell's work.

In quantum mechanics, if we consider three compatible propositions, \{A, B,
C\}, (associated to pairwise commuting projectors) the inequalities $\left( 
\ref{26}\right) $ hold true because the lattice of commuting projectors is
distributive. On the other hand, if two of the propositions, say A and B,
are not compatible then their distance is not defined because quantum
mechanics does not provide a joint probability of two incompatible
observables (and it is assumed that they cannot be measured simultaneously).
However there are cuadrilateral inequalities, derived from the triangular
ones $\left( \ref{26}\right) ,$ which may be violated by quantum mechanics
and tested empirically. In fact, if we consider four projectors \{A, B, C,
D\} it is easy to see that the inequalities $\left( \ref{26}\right) $ lead
to 
\begin{equation}
d(A,D)\leq d(A,B)+d(B,C)+d(C,D).  \label{27}
\end{equation}
At a diference with the inequalites $\left( \ref{26}\right) $, now all four
distances are defined in quantum mechanics if every pair involve commuting
projectors (that is if $\left[ A,D\right] =\left[ A,B\right] =\left[
B,C\right] =\left[ C,D\right] =0$). We see that the inequality $\left( \ref
{27}\right) $and the other three obtained by permutations involving the four
projectors are necessary conditions for the existence of a \textit{classical}
joint probability distribution defined on the set of projectors. The are
cases where quantum mechancis predicts violations of one ot the
inequalities, which leads to Bell's theorem (see next section.)

Inequality $\left( \ref{27}\right) $ is equivalent to the following one 
\begin{equation}
p_{B}+p_{C}\geq p_{AB}+p_{BC}+p_{CD}-p_{DA},  \label{8}
\end{equation}
where p$_{A}$ (or p$_{AB}$) is the probability that $A$ (or $AB$) is true.
This is called a Bell inequality \cite{B2} and, in this form it was derived
by Clauser and Horne \cite{CH}. Instead of projectors, taking the values 0
or 1, we might use observables taking the values -1 or +1. They are
trivially related to the projectors by 
\begin{equation}
a=2A-1,b=2B-1,etc.  \label{9}
\end{equation}
and the inequality $\left( \ref{8}\right) $ takes the form of
Clauser-Horne-Shimony-Holt (CHSH) \cite{CHSH}: 
\begin{equation}
\left| \left\langle ab\right\rangle +\left\langle bc\right\rangle
+\left\langle cd\right\rangle -\left\langle ad\right\rangle \right| \leq 2,
\label{10}
\end{equation}
where $\left\langle ab\right\rangle $ means the expectation value of the
product of \textit{a} and \textit{b}. Therefore these, CHSH, and the
Clauser-Horne inequalities $\left( \ref{8}\right) $ are equivalent.

\section{Hidden variables theories}

The question of hidden variables in quantum mechanics aroused soon after the
formulation of the theory during the years 1925-26. It was explicitly
mentioned in the book by von Neumann in 1932 \cite{vN}, where he derived a
celebrated no hidden variables theorem. From that time many books and
articles have been devoted to the subject. Nevertheless there is no sharp
definition of hidden variables (HV) theory which is widely accepted. I
propose the following:

\begin{definition}
HV is a theory physically equivalent to quantum mechanics (that is giving
the same predictions for all experiments) which has the formal structure of
classical statistical mechanics.
\end{definition}

The definition may be illustrated in the following table giving the
correspondence of concepts in experiments, standard quantum theory and a
possible HV theory:

\medskip

\textbf{Table I. Correspondence of concepts}

\medskip

$
\begin{array}{lll}
\text{EMPIRICAL} & \text{QUANTUM THEORY} & \text{HV THEORY} \\ 
\text{physical system} & \text{Hilbert space H} & \text{phase space }\Lambda
\\ 
\text{state} & \text{vector }\mid \Psi \rangle \in \text{ H} & \text{%
probability density }\rho \left( \lambda \right) \\ 
\text{observable A} & \text{self-adjoint operator }\widehat{A} & \text{%
function }A\left( \lambda \right) \\ 
\text{expectation value} & \langle \Psi \mid \widehat{A}\mid \Psi \rangle & 
\text{= }\int A\left( \lambda \right) \rho \left( \lambda \right) d\lambda
\\ 
\text{correlation} & \text{if }\widehat{A}\text{ }\widehat{B}\text{ = }%
\widehat{B}\text{ }\widehat{A}\text{, }\langle \Psi \mid \widehat{A}\text{ }%
\widehat{B}\mid \Psi \rangle & =\text{ }\int A\left( \lambda \right) B\left(
\lambda \right) \rho \left( \lambda \right) d\lambda
\end{array}
$

\smallskip

\medskip

The parameter\ (or parameters) $\lambda $ is usually called the \textit{%
hidden variable}. Two observables, A and B, which are associated to
commuting operators, $\widehat{A}$ and $\widehat{B}$, are said \textit{%
compatible. }The correlation may be extended to more than two compatible
observables. It is easy to see that the latter equality implies the equality
of the joint probability distributions of compatible observables. In fact,
it is enough to substitute exp$\left( i\xi \widehat{A}\right) $ for $%
\widehat{A}$ and exp$\left[ i\xi A\left( \lambda \right) \right] $ for $%
A\left( \lambda \right) $ in the equality, and similarly for $B$, in order
to show the equality of the characteristic function of the joint probability
distribution. On the other hand, it is well known that quantum mechanics
does not provide joint distributions of observables not compatible (the
associated operators noncommuting). For the sake of clarity, in the Table we
have considered only quantum pure states. The most general states are
associated to density operators, $\widehat{\rho }$, whence the quantum
expectation value and correlation should be written, respectively 
\[
Tr\left( \widehat{\rho }\widehat{A}\right) ,Tr\left( \widehat{\rho }\widehat{%
A}\widehat{B}\right) . 
\]

In order to make clear what is the content of the theorems against HV
theories, discussed later, I propose the following

\begin{definition}
A simple experiment consists of the preparation of a state of a physical
system, followed by the evolution of the system and finishing with the
measurement of a set of compatible observables.
\end{definition}

\begin{definition}
A composite experiment consists of several simple experiments with the same
preparation and the same subsequent evolution, but measuring different sets
of compatible observables in each simple experiment.
\end{definition}

With these definitions we may state the following theorem:

\begin{theorem}
For any simple experiment there exists a HV theory.
\end{theorem}

\textit{Proof:} The essential part of the proof is to show that for any
state $\mid \Psi \rangle $ and two compatible observables $A,$ $B$ the
expectation may be obtained in the form 
\begin{equation}
\langle \Psi \mid \widehat{A}\widehat{B}\mid \Psi \rangle =\int A\left(
\lambda \right) B\left( \lambda \right) \rho \left( \lambda \right) d\lambda
.  \label{1}
\end{equation}
For simplicity we consider just two observables, but the generalization to
any finite number is trivial. In order to proceed with the proof we recall
that there exists a complete set of orthonormal vectors which are
simultaneous eigenvectors of two commuting self-adjoint operators. Let us
label $\mid \lambda \rangle $ one of the common eigenvectors of $\widehat{A}$
and $\widehat{B}$. Complete means that 
\begin{equation}
\int \mid \lambda \rangle \langle \lambda \mid d\lambda =1,  \label{2}
\end{equation}
which leads to 
\begin{eqnarray}
\langle \Psi \left| \widehat{A}\widehat{B}\right| \Psi \rangle &=&\int
d\lambda d\lambda ^{\prime }d\lambda ^{\prime \prime }\langle \Psi \left|
\lambda \rangle \langle \lambda \right| \widehat{A}\left| \lambda ^{\prime
}\rangle \langle \lambda ^{\prime }\right| \widehat{B}\left| \lambda
^{\prime \prime }\rangle \langle \lambda ^{\prime \prime }\right| \Psi
\rangle  \nonumber \\
&=&\int d\lambda \langle \Psi \left| \lambda \rangle \langle \lambda \right| 
\widehat{A}\left| \lambda \rangle \langle \lambda \right| \widehat{B}\left|
\lambda \rangle \langle \lambda \right| \Psi \rangle  \label{3a} \\
&=&\int d\lambda \left| \langle \Psi \mid \lambda \rangle \right|
^{2}\langle \lambda \left| \widehat{A}\right| \lambda \rangle \langle
\lambda \left| \widehat{B}\right| \lambda \rangle .\smallskip  \label{3}
\end{eqnarray}
This has the structure of the right side of eq.$\left( \ref{1}\right) $
provided we identify $\langle \lambda \mid \widehat{A}\mid \lambda \rangle $
with the function $A\left( \lambda \right) $ and $\left| \langle \Psi \mid
\lambda \rangle \right| ^{2}$ with the density $\rho \left( \lambda \right)
. $ Indeed the density is positive and normalized (the latter because eq.$%
\left( \ref{2}\right) ).$ Eq.$\left( \ref{3a}\right) $ follows from the
equality 
\begin{equation}
\langle \lambda \mid \widehat{A}\mid \lambda ^{\prime }\rangle =\langle
\lambda \mid \widehat{A}\mid \lambda \rangle \delta \left( \lambda -\lambda
^{\prime }\right) ,  \label{4}
\end{equation}
$\delta $ being Dirac's delta, which is a consequence of $\mid \lambda
\rangle $ and $\mid \lambda ^{\prime }\rangle $ being eigenvectors of $%
\widehat{A}.$

We see that hidden variables are always possible, a fact made clair by J. S.
Bell in 1966 \cite{B1}. However, some families of HV theories are excluded,
for instance those in which expectations fulfil linear relations of the form 
\begin{equation}
\langle \Psi \mid \widehat{A}+\widehat{B}\mid \Psi \rangle =\int \left[
A\left( \lambda \right) +B\left( \lambda \right) \right] \rho \left( \lambda
\right) d\lambda ,  \label{5}
\end{equation}
The impossibility of such HV\ theories is the content of von Neumann's
theorem mentioned above \cite{vN}. Assumption $\left( \ref{5}\right) $ is
unphysical, as pointed out by Bell \cite{B1}, which shows that von
Neumann\'{}s theorem is not very relevant. More physical requirements are
noncontextuality and locality which we discuss in the following.

\section{Noncontextual hidden variables}

\begin{definition}
A HV theory is noncontextual if there exists a joint probability
distribution for all observables of the system (even if some of them are not
compatible.)
\end{definition}

In particular this implies that the marginal for the variable A in the joint
distribution of the compatible observables A and B is the same as the
marginal for A in the joint distribution of the compatible observables A and
C, even if B and C are not compatible. For this reason noncontextuality is
sometimes stated saying that the result of measuring A does not depend on
the context (in particular, the result is the same whether we measure A
toghether with B or we measure A toghether with C; remember that A, B, C
cannot be measured simultaneously, that is with the same experimental set
up). The latter property is true in quantum mechanics, but the existence of
a joint distribution is a stronger constraint. What is required is the
existence of some function of all the observables, p(A,B,C...), which
fulfils the mathematical properties of a joint probability distribution and
it is such that the marginals for every subset of compatible observables is
the same given by quantum mechanics. The said distribution is just a
mathematical object (it cannot be measured if some of the observables are
not compatible) but their mere existence puts constraints which may be
tested empirically.

It is not difficult to see that the existence of a joint distribution for
the observables A, B, C,... is equivalent to the existence of a positive
normalized function, $\rho \left( \lambda \right) $ of a variable or set of
variables, $\lambda ,$ and functions A$\left( \lambda \right) ,$ B$\left(
\lambda \right) ,$ C$\left( \lambda \right) $ ... However a joint
probability distribution cannot be obtained with the construction of eq.$%
\left( \ref{3}\right) $ if the observables are not compatible$.$ This is
because a complete orthonormal set of simultaneous eigenvectors of $\widehat{%
A}$, $\widehat{B}$, $\widehat{C}$,... may not exist if the operators do not
commute pairwise. What may be obtained are several HV theories, one for each
simple experiment. For instance, let us consider a composite experiment
consisting of two simple ones. In the first, where we measure A and B, a HV
theory should provide the functions $\rho _{1}\left( \lambda \right) $, A$%
_{1}\left( \lambda \right) ,$ B$_{1}\left( \lambda \right) $. In the second,
where we measure A and C, a\ HV theory would give $\rho _{2}\left( \lambda
\right) $, A$_{2}\left( \lambda \right) ,$ C$_{2}\left( \lambda \right) $ .
The two HV theories toghether might be called a HV theory for the composite
experiment. It would be noncontextual if $\rho _{1}\left( \lambda \right) $
= $\rho _{2}\left( \lambda \right) $ and A$_{1}\left( \lambda \right) $ = A$%
_{2}\left( \lambda \right) ,$ if this does not happen it should be \textit{%
contextual.}

The impossibility of noncontextual theories is stablished by the following

\begin{theorem}
Noncontextual HV theories do not exist for all (composite) experiments.
\end{theorem}

This is usually called Kochen-Specker theorem \cite{KS} after the authors
who proved it in 1967. However the theorem had been actually proved one year
earlier by Bell \cite{B1} and it is a rather direct consequence of a theorem
proved in 1957 by Gleason \cite{G}. We shall give here a proof inspired in
the celebrated theorem of Bell against local hidden variables \cite{B2}.

\textit{Proof}: It is enough to exhibit a particular type of composite
experiment where the quantum predictions are incompatible with the existence
of a joint probability distribution for all observables. We consider four
dichotomic observables, A, B, C and D, each of which may take the values 0
or 1. We assume that A and C are not compatible, and B and D are also not
compatible, the remaining pairs being compatible. The corresponding
operators will be proyectors, i. e. $\widehat{A}^{2}$ = $\widehat{A},$ etc.,
all pairs commuting except 
\begin{equation}
\left[ \widehat{A},\widehat{C}\right] \neq 0,\left[ \widehat{B},\widehat{D}%
\right] \neq 0.  \label{6}
\end{equation}
Let us label p$_{A}$ the probability of A = 1, p$_{AB}$ the probability that
A = B = 1, etc. The existence of a joint distribution means that there are
15 positive quantities 
\begin{equation}
p_{A},p_{B},p_{C},p_{D},p_{AB},p_{AC},p_{AD},p_{BC},p_{BD},p_{CD},p_{ABC},p_{ABD},p_{ACD},p_{BCD},p_{ABCD},
\label{6a}
\end{equation}
which should fulfil the relations 
\begin{equation}
0\leq p_{ABCD}\leq p_{ABC}\leq p_{AB}\leq p_{A}\leq 1,  \label{7}
\end{equation}
and those obtained by all permutations of the labels. Only 8 of these
quantities may be measured (and they are predicted by quantum mechanics),
namely 
\begin{equation}
p_{A},p_{B},p_{C},p_{D},p_{AB},p_{AD},p_{BC},p_{CD}.  \label{7a}
\end{equation}
The remaining 7 quantities cannot be measured, the corresponding observables
not being compatible, and quantum mechanics gives no value for them.

The question is whether there exist 7 quantities fulfilling all constraints
of the type $\left( \ref{7}\right) $ which added to the 8 measurable ones
provide the desired joint probability distribution $\left( \ref{6a}\right) $%
. Now a necessary condition for the existence of a joint probability
distribution is the fulfillement of the Bell inequalities discussed above. A
sufficient condition is the fulfillement of the 4 Bell inequalities obtained
by suitable permutation of labels in $\left( \ref{8}\right) $ or $\left( \ref
{10}\right) $ (that is substituting A for C or D for B or both) \cite{F}.
The rest of the proof consists of showing that there are states and
observables for which quantum mechanics violates the inequalities, which may
be seen elsewhere, e. g. \cite{CH}.

\section{Local hidden variables}

An important class of HV theories are \textit{local HV theories. }The
concept of local\textit{\ }applies to EPR experiments. We call EPR \cite{EPR}
an experiment where we prepare locally a system which is later divided in
two subsystems, each of which moves in a different direction. Measurements
on each subsystem are later made at space-like separation (in the sense of
relativity theory).

\begin{definition}
\smallskip A HV theory is local if, for any EPR experiment where we may
measure one of several observables, A$_{i}$, of the first subsystem and one
of several observables, B$_{j}$, on the second, there exist a joint
probability distribution for all the observables $\left\{
A_{i},B_{j};i,j=1,2,....\right\} .$
\end{definition}

The impossibility of local HV theories is stablished by the celebrated
Bell's theorem of 1964 \cite{B2}

\begin{theorem}
Local HV theories do not exist for all (EPR) experiments.
\end{theorem}

\textit{Proof}: The proof is the same as for noncontextual HV theories, but
considering an EPR experiment. That is, the observables A, C belong to one
subsystem and B, D to the other subsystem. In particular, this guarantees
that the pairs \{A,\ B\}, \{A,\ D\}, \{C,\ B\}, \{C,\ D\} are compatible
because they belong to spacelike separated regions (the condition that
spacelike separated observables are compatible is called \textit{%
microcausality} in quantum field theory).

The class of local theories is wider than that of noncontextual HV theories
because the constraints in their definition are weaker. Indeed in local
theories the existence of a joint distribution is only required for EPR
experiments, but noncontextual theories assume it for all experiments.
Consequently the empirical disproof is easier for noncontextual theories
than for local theories. In the former it is enough to perform a composite
experiment where the measurments are made locally, the latter requires
measurements at spacelike separation.

The fact that the proofs of both theorems are very similar has been a source
of misundrstanding, like the assertion that locality is not needed in order
to prove Bell's theorem. I hope that in our presentation the point is more
clear. But in order to stress the distinction between noncontextual and
local I give an illustrative example.

Let us assume that we want to perform a test of Bell\'{}s inequality using
two spin-1/2 particles prepared in a singlet (zero total spin) state. An
example could be the dissociation of a molecule consisting of two sodium
atoms. We should measure the spin components along two directions for each
atom (in four different\textit{\ simple} experiments, see section 3 for the
definition of simple experiment). These directions define the four
projectors involved in the Bell inequality. If the inequality is violated we
would have an empirical disproof of noncontextual hidden variables theories.
However if we want to test local theories, the measurements should be
performed at space-like separation, which is a rather strong requirement.

For instance, we might use two Stern-Gerlach apparatuses each of length L.
If the atoms move at velocity v, in opposite directions, the duration of the
measurement would be L/v. The condition that the measurements are space-like
separated means that the distance, d, between the Stern-Gerlach apparatuses
should fulfil 
\[
d>2L\frac{c}{v}. 
\]
This inequality involves the velocity of light as it should, locality (in
the sense of Bell) being a relativistic concept. An estimate of the minimal
distance is obtained if we use a typical energy involved in dissociation,
say 1 eV, and L is of the order of a few centimeters. We get for the minimal
distance several kilometers. Thus the empirical violation of local hidden
variables is far more difficult than the violation of non-contextual ones.

\section{Quantum information}

The amount of information is quantified with the concept of \textit{entropy}%
. In classical physics, if we have a continuous random variable, $\lambda ,$
with a probability distribution $\rho \left( \lambda \right) ,$ the entropy,
S$^{C}$ , as defined by Shannon is 
\begin{equation}
S^{C}=-\int \rho \left( \lambda \right) \log \rho \left( \lambda \right)
d\lambda .  \label{11}
\end{equation}
The quantum entropy was defined by von Neumann in terms of the density
operator, $\widehat{\rho }$, with an expression which looks similar to that
one, namely 
\begin{equation}
S^{Q}=-Tr\left( \widehat{\rho }\log \widehat{\rho }\right) .  \label{12}
\end{equation}
In both cases S $\geq 0$ and the entropy increases with the lack of
information, so that the pure states (maximal information) corresponds to S
= 0.

There are two other properties which hold true for both classical and
quantum entropy:

\textbf{Concavity: }$\lambda S\left( \rho _{a}\right) +(1-\lambda )S\left(
\rho _{b}\right) \leq S\left( \lambda \rho _{a}+(1-\lambda )\rho _{b}\right)
,0\leq \lambda \leq 1,$

where $\rho _{a}$ stands for either the classical probability density, $\rho
_{a}\left( \lambda \right) ,$ or the quantum density operator, $\widehat{%
\rho }_{a}$ and similarly $\rho _{b}$ for a different probabilty density or
density operator of the same system.

\textbf{Subadditivity}: $S\left( \rho _{12}\right) \leq S\left( \rho
_{1}\right) +S\left( \rho _{1}\right) ,$

where $\rho _{12}$ stands for either the classical probability density, $%
\rho _{12}\left( \lambda _{1},\lambda _{2}\right) ,$ or the quantum density
operator, $\widehat{\rho }_{12}$ , the subindex 1 (2) referring to the first
(second) subsystem of a composite system, and we have 
\begin{equation}
\rho _{1}\left( \lambda _{1}\right) =\int \rho _{12}\left( \lambda
_{1},\lambda _{2}\right) d\lambda _{2},\widehat{\rho }_{1}=Tr_{_{2}}\widehat{%
\rho }_{12}.  \label{13}
\end{equation}

There is, however, a property which dramatically distinguish classical from
quantum entropy. In fact in the case of a system consisting of two
subsystems, the classical, Shannon's, entropy fulfils 
\begin{equation}
S^{C}\left( \rho _{12}\right) \geq \max \left\{ S^{C}\left( \rho _{1}\right)
,S^{C}\left( \rho _{2}\right) \right\} ,  \label{14}
\end{equation}
whilst the quantum entropy fulfils the weaker triangle inequality 
\begin{equation}
S^{Q}\left( \widehat{\rho }_{12}\right) \geq \left| S^{Q}\left( \widehat{%
\rho }_{1}\right) -S^{Q}\left( \widehat{\rho }_{2}\right) \right| .
\label{15}
\end{equation}

In my opinion, the fact that the quantum entropy does not fulfil an
inequality similar to $\left( \ref{14}\right) $ is highly paradoxical, I
would even say bizarre. In fact, $\left( \ref{15}\right) $ allows for the
possibility that both $S^{Q}\left( \widehat{\rho }_{1}\right) $ and $%
S^{Q}\left( \widehat{\rho }_{2}\right) $ are positive whilst $S^{Q}\left( 
\widehat{\rho }_{12}\right) $ is zero. This should be interpreted saying
that we have complete information about a composite system whilst we have
incomplete information about every subsystem. This contrast with the
classical, and intuitive, idea that full information about the whole \textit{%
means }that we have complete information about every part. In my view this
is indicative that the concept of ''complete'' information in quantum theory
is not the same as in classical physics, and the different meanings of
completeness has been the source of misunderstandings about the
interpretation of quantum theory, e.g. in the debate between Einstein and
Bohr.

The violation of an inequality similar to $\left( \ref{14}\right) $ is
closely related to the violation of the Bell inequality. But in order to
stablish the conection it is necessary to introduce the concept of \textit{%
linear entropy.} Actually, although the definitions of entropy $\left( \ref
{11}\right) $ and $\left( \ref{12}\right) $ are standard and in some sense
an optimum, it is possible to give alternative definitions of entropy which
fulfil the essential properties of concavity and subadditivity. The most
simple are the socalled linear entropies 
\begin{equation}
S^{CL}=1-\int \rho \left( \lambda \right) ^{2}d\lambda ,S^{QL}=1-Tr\left( 
\widehat{\rho }^{2}\right) .  \label{16}
\end{equation}

The desired connection between linear entropy and the Bell inequalities has
been studied by several authors in the last few years. For instance
Horodecki et al. \cite{HH} proved that the inequality $\left( \ref{14}%
\right) $ is a sufficient condition for the Bell inequalities. A slightly
stronger result may be stated as follows

\begin{theorem}
The inequality 
\begin{equation}
MNS^{QL}\left( \widehat{\rho }_{12}\right) +MN-M-N\geq NS^{QL}\left( 
\widehat{\rho }_{1}\right) +MS^{QL}\left( \widehat{\rho }_{2}\right) ,
\label{17}
\end{equation}
where M and N are the dimensions of the Hilbert spaces of the two
subsystems, is a sufficient condition for all Bell inequalities $\left( \ref
{8}\right) $ or $\left( \ref{10}\right) $ which may be got using two
dichotomic observables of each subsystem.
\end{theorem}

\textit{Proof: }We consider observables \{a, b\} for the first particle and
\{c, d\} for the second, all of which may take values 1 or -1, and the
associated operators, $\widehat{a}$, $\widehat{b}$ , $\widehat{c}$ and $%
\widehat{d}.$ We define the Bell operator, $\widehat{B}$ , by 
\begin{equation}
\widehat{B}=\widehat{a}\otimes \widehat{b}+\widehat{c}\otimes \widehat{b}+%
\widehat{c}\otimes \widehat{d}-\widehat{a}\otimes \widehat{d.}  \label{18}
\end{equation}
It is easy to see that 
\begin{equation}
Tr\widehat{B}=0,Tr\left( \widehat{B}^{2}\right) =4MN,  \label{19}
\end{equation}
and that the Bell inequality $\left( \ref{10}\right) $ is violated if 
\begin{equation}
\left| \beta \right| >2,\beta \equiv Tr\left( \widehat{B}\widehat{\rho }%
_{12}\right) ,  \label{20}
\end{equation}
(whilst quantum mechanics predicts just $\left| \beta \right| \leq 2\sqrt{2}%
).$ Now the obvious inequality 
\begin{equation}
Tr\left( \widehat{\rho }_{12}-\frac{1}{N}\widehat{\rho }_{1}\otimes \widehat{%
I_{2}}-\frac{1}{M}\widehat{I_{1}}\otimes \widehat{\rho }_{2}+\frac{1}{MN}%
\widehat{I_{1}}\otimes \widehat{I_{2}}+\lambda \widehat{B}\right) ^{2}\geq
0,\lambda \in R,  \label{21}
\end{equation}
where $\widehat{I_{1}}($ $\widehat{I_{2}}$ ) is the identity operator for
the first (second) particle, gives a quadratic expression in the variable $%
\lambda .$ We get, after some algebra 
\begin{equation}
MNTr\left( \widehat{\rho }_{12}^{2}\right) -NTr\left( \widehat{\rho }%
_{1}^{2}\right) -MTr\left( \widehat{\rho }_{2}^{2}\right) \geq \frac{1}{4}%
\left( \beta ^{2}-4\right) .  \label{22}
\end{equation}
Hence the inequality $\left( \ref{17}\right) $ implies $\left| \beta \right|
\leq 2$ $,$ which proves the theorem.

Actually the inequality $\left( \ref{17}\right) $ is rather strong, and
therefore not very useful, if either M \TEXTsymbol{>} 2 or N \TEXTsymbol{>}
2 or both, and it is trivial if either M = 1 or N = 1. Consequently its main
interest is the case M\ = N = 2, where it is a consequence of the inequality 
$\left( \ref{14}\right) $ characteristic of classical information theory.

\textbf{Acknowledment. }I acknowledge financial support from DGICYT, Project
No. PB-98-0191 (Spain).

\end{document}